\documentclass[aps,prd,floatfix,twocolumn,superscriptaddress,preprintnumbers,nofootinbib]{revtex4-1}

\usepackage{textcomp} 
\usepackage{slashed}
\usepackage{epsfig,latexsym,cancel,amssymb,amsmath,verbatim,mathrsfs}
\usepackage{color}
\usepackage{graphicx}
\usepackage{enumitem}
\usepackage[colorlinks,citecolor=blue]{hyperref}

\newcommand{\beq}{\begin{equation}}
\newcommand{\eeq}{\end{equation}}
\newcommand{\bea}{\begin{eqnarray}}
\newcommand{\eea}{\end{eqnarray}}
\newcommand{\nn}{\nonumber}

\usepackage{ulem}

\allowdisplaybreaks[1]

\begin{document}

\preprint{
	{\vbox {
			\hbox{\bf LA-UR-22-20027}
			\hbox{\bf MSUHEP-22-001}
}}}
\vspace*{0.2cm}

\title{Probing the $Zb\bar{b}$ anomalous couplings via exclusive $Z$ boson decay}

\author{Hongxin Dong}
\email{211002010@njnu.edu.cn}
\affiliation{Department of Physics and Institute of Theoretical Physics, Nanjing Normal University, Nanjing, Jiangsu 210023, China}

\author{Peng Sun}
\email{sunpeng@njnu.edu.cn}
\affiliation{Department of Physics and Institute of Theoretical Physics, Nanjing Normal University, Nanjing, Jiangsu 210023, China}

\author{Bin Yan}
\email{binyan@lanl.gov}
\affiliation{Theoretical Division, Group T-2, MS B283, Los Alamos National Laboratory, P.O. Box 1663, Los Alamos, NM 87545, USA}

\author{C.-P. Yuan}
\email{yuan@pa.msu.edu}
\affiliation{Department of Physics and Astronomy,
Michigan State University, East Lansing, MI 48824, USA}

\begin{abstract}
We propose to utilize the exclusive $Z$-boson rare decays $Z\to \Upsilon(ns)+\gamma$ to constrain the $Zb\bar{b}$ couplings at the HL-LHC and 100 TeV proton-proton collider. We demonstrate that the event yield of the proposed  processes is sensitive to the axial-vector component of the $Zb\bar{b}$ coupling and can provide
complementary information to the jet-charge weighted single-spin asymmetry measurement at the EIC and the $gg\to Zh$ production rate measurement at the LHC.
By applying the NRQCD factorization formalism, we calculate the partial decay width of $Z\to \Upsilon(ns)+\gamma$ to the  NLO accuracy in strong interaction, which is found to agree with those obtained from the light-cone distribution amplitude approach. 
We show that  the HL-LHC can break the degeneracy of the $Zb\bar{b}$ couplings, as implied by the precision electroweak data at LEP and SLC, if the signal  efficiency can be improved by a factor of 1.7 from the present ATLAS analysis at the 13 TeV LHC with an integrated luminosity of $36.1~{\rm fb}^{-1}$.
\end{abstract}

\maketitle

\section{Introduction}
The discovery of the Higgs boson at the Large Hadron Collider (LHC) marked the observation of the full spectrum of elementary particles predicted by the Standard Model (SM). Despite the great success of the SM, there are several aspects of nature for which the
SM description is completely lacking or unsatisfactory, which motivate the search for new physics (NP) beyond the SM either directly at the
LHC or indirectly with lower energy, high precision experiments.
The electroweak precision measurements at the LEP and SLC have received much attention in the high energy physics community, due to the remarkable accuracy of the data, and have imposed strong constraints on various NP models~\cite{Wells:2014pga,Berthier:2015oma}. It is evident that one of the most interesting electroweak measurements  is the bottom quark forward-backward asymmetry ($A_{\rm FB}^b$) at the $Z$-pole from the LEP, which exhibits a long-standing discrepancy with the SM prediction, with a significance about $2.1\sigma$~\cite{ParticleDataGroup:2020ssz}. Such anomaly could be translated into the deviation of the  $Zb\bar{b}$ couplings from the SM prediction. As shown in Refs.~\cite{Choudhury:2001hs,Gori:2015nqa}, a large deviation of the right-handed $Zb\bar{b}$ coupling, with a SM-like left-handed $Zb\bar{b}$ coupling, is needed to explain the $A_{\rm FB}^b$ data, together with the measurements of the branching fraction ($R_b$) of $Z\to b\bar{b}$ in the inclusive hadronic decay of $Z$ boson at the LEP and the bottom quark left-right forward-backward asymmetry ($A_b$) at the SLC. Such a condition can be fulfilled for any theory with an underlying approximate custodial symmetry~\cite{Agashe:2006at} and has been widely discussed in the literature~\cite{Choudhury:2001hs,Agashe:2006at,Gori:2015nqa,Liu:2017xmc}. 
It is also well know that the $Zb\bar{b}$ anomalous couplings are not fully determined by the electroweak precision measurements since the appearance of the  degeneracy under the global analysis~\cite{Choudhury:2001hs}. Recently, several approaches have been proposed in the literature to break the  above mentioned degeneracy and to further test the $Zb\bar{b}$ couplings at the LHC and future colliders~\cite{Yan:2021veo,Yan:2021htf,Li:2021uww}. For example, the axial-vector component of the $Zb\bar{b}$ coupling can be determined by the precision measurement of the $gg\to Zh$ production at the LHC and high-luminosity LHC (HL-LHC), a proton-proton collider to operate at a center-of-mass energy of 14 TeV with an integrated luminosity of $3~{\rm ab}^{-1}$~\cite{Yan:2021veo}. It can also be better constrained by the average jet charge weighted single-spin asymmetry to be measured at the upcoming Electron-Ion Collider (EIC)~\cite{Li:2021uww}, while the  vector-component of the $Zb\bar{b}$ coupling can be determined by the measurement of the single-spin asymmetry, of the polarized
electron-proton cross section in neutral current deeply inelastic
scattering processes with one $b$-tagged jet in the final state, at the HERA and the EIC~\cite{Yan:2021htf}. 

In this work, we propose yet another novel idea to probe the axial-vector component of the $Zb\bar{b}$ coupling at the HL-LHC and 100 TeV proton-proton (pp) collier through the exclusive $Z$-boson decay process  $Z\to \Upsilon(ns)+\gamma$, with $n=1,2,3$. Because the $J^{PC}$ quantum number of $\Upsilon(ns)$ and $\gamma$ are $J^{PC}=1^{--}$, the vector-component of the $Zb\bar{b}$ coupling can not contribute to this rare decay process due to the charge conjugation symmetry or Furry's theorem. 
This unique feature, together with 
the large event rate of the inclusive $Z$ boson production and the easily identifiable final state of the decay process $\Upsilon(ns)\to \ell^+\ell^-$, with $\ell^{\pm}=e^{\pm},\mu^\pm,\tau^\pm$, makes $Z\to \Upsilon(ns)+\gamma$ an ideal channel to directly probe the axial-vector component of the $Zb\bar{b}$ coupling  at hadron colliders.   

The exclusive $Z$-boson decay process $Z\to \Upsilon(ns)+\gamma$ has been widely discussed in the framework of non-relativistic QCD (NRQCD), at the leading order (LO) in strong coupling $\alpha_s$~\cite{Guberina:1980dc,Huang:2014cxa}, or using the light-cone distribution amplitude (LCDA) approach~\cite{Huang:2014cxa,Grossman:2015cak,Bodwin:2017pzj}. In this paper, we will consider the next-to-leading order (NLO) QCD correction for the decay width of $Z\to \Upsilon(ns)+\gamma$ in the NRQCD framework. We will demonstrate below that this rare decay process is indeed dependent only on the axial-vector component of the $Zb\bar{b}$ coupling, and can help to  determine the $Zb\bar{b}$ coupling at the HL-LHC and 100 TeV pp colliders.
 
\section{Theoretical analysis}
The exclusive decay width of $Z\to \Upsilon(ns)+\gamma$, in the framework of NRQCD, can be written as~\cite{Bodwin:1994jh},
\beq
\Gamma[Z\to \Upsilon(ns)+\gamma]=\hat{\Gamma}[Z\to (b\bar{b})+\gamma]\langle\mathcal{O}^{\Upsilon(ns)}(^3S_1)\rangle.
\label{eq:width}
\eeq
Here, $\hat{\Gamma}[Z\to (b\bar{b})+\gamma]$ is the short-distance coefficient, which is independent of the bottomonium state and can be obtained  by matching the calculation of perturbative QCD and NRQCD. The effect of non-perturbative physics is described by the long-distance matrix element $\langle\mathcal{O}^{\Upsilon(ns)}(^3S_1)\rangle$, whose value can be extracted  from the experimental measurement of the decay width $\Gamma[\Upsilon(ns)\to e^+e^-]$.  We note that in the framework of NRQCD, the relativistic corrections to this decay width will be suppressed by $\mathcal{O}(v^2)$, with $v$ being the relative velocity of the bottom quarks in the meson rest frame. It was shown in Refs.~\cite{Huang:2014cxa,Bodwin:2017pzj} that its numerical effect is very small and will be ignored in this work.  
Below, we will calculate the partial decay width of $\Gamma[Z\to \Upsilon(ns)+\gamma]$ at the LO and NLO, using the NRQCD factorization formalism.
In order to consider the impact of the non-standard $Zb\bar{b}$ couplings to the exclusive radiative decay  $Z\to \Upsilon(ns)+\gamma$, we consider the following effective Lagrangian:
\beq
\mathcal{L}_{\rm eff}=\frac{g_W}{2c_W}\bar{b}\gamma_\mu\left(\kappa_V g_V^b-\kappa_A g_A^b\gamma_5\right)b Z_\mu,
\label{eq:effL}
\eeq
where $g_W$ is the $SU(2)_L$ gauge coupling. The parameters $g_V^b=-1/2+2/3s_W^2$ and $g_A^b=-1/2$ are the vector and axial-vector components of the $Zb\bar{b}$ coupling in the SM, respectively. Here $c_W=\cos\theta_W$ and $s_W=\sin\theta_W$, with $\theta_W$ being the weak mixing angle of the SM. The parameters $\kappa_{V,A}$ are introduced to parametrize possible NP effects and $\kappa_{V,A}=1$ in the SM.
Although the dipole operators (with $\sigma_{\mu\nu}$ term) could also contribute to the anomalous $Z-b-{\bar b}$ couplings, its contribution to the exclusive decays $Z \to \Upsilon(ns)+\gamma$ would be a sub-leading effect as compared to that from $\kappa_{V,A}$. This is because the dipole operators can only be induced at loop level~\cite{Arzt:1994gp}, while $\kappa_{V,A}$ can be generated at tree level (see, for example, Ref.~\cite{Choudhury:2001hs}) from a renormalizable ultraviolet (UV) completion theory. In this study, we only consider the impact of new physics effects which can be parametrized in the form of $\kappa_{V,A}$, as shown in Eq.~\eqref{eq:effL}.
In the Standard Model effective field theory (SMEFT) context, the anomalous  
$\kappa_{V,A}$ can be matched to some dimension-6 effective operators after the electroweak symmetry breaking~\cite{Buchmuller:1985jz,Grzadkowski:2010es}. 
Though the dimension-6 dipole operators could also contribute to the rare decay processes under consideration, their effect will be ignored in this study.

\subsection{The LO decay rate}
\begin{figure}
\centering
\includegraphics[scale=0.4]{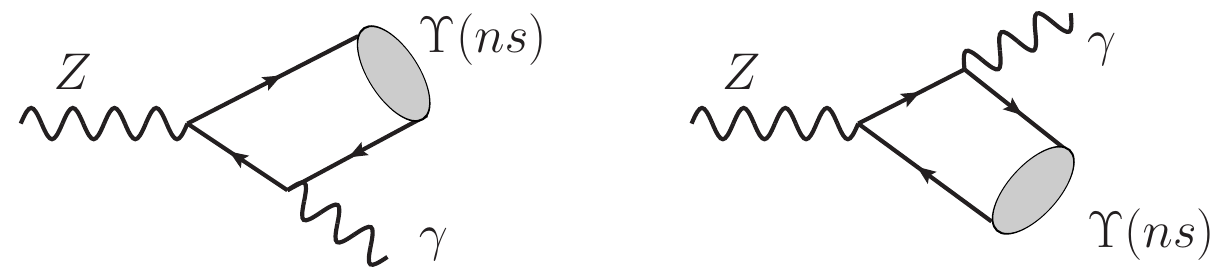}
\caption{The leading order Feynman diagrams of $Z\to \Upsilon(ns)+\gamma$. }
\label{Fig:FeyLO}
\end{figure}

At the LO, there are two Feynman diagrams which can contribute to the exclusive radiative decay $Z(p_Z)\to \Upsilon(ns)(2p_b)+\gamma (p_\gamma)$; see Fig.~\ref{Fig:FeyLO}.
The amplitude can be calculated by using the covariant projection operator, which is defined as
\beq
\Pi=\frac{\Psi_{\Upsilon(ns)}(0)}{2\sqrt{m_\Upsilon}}\slashed{\epsilon}_\Upsilon^*(p_\Upsilon)\left(\slashed{p}_{\Upsilon}+m_{\Upsilon}\right)\otimes\frac{\bold{1}_c}{\sqrt{N_c}},
\eeq
where, $N_c=3$ and $\bold{1}_c$ denotes the unit color matrix.
$\epsilon_\Upsilon^\mu(p_\Upsilon)$ is the polarization vector of the $\Upsilon$ with the momentum $p_\Upsilon$, and  $\Psi_{\Upsilon(ns)}(0)$ is the Schr\"odinger wave function of the $\Upsilon(ns)$ at the origin.
In the framework of NRQCD, we have $p_\Upsilon=2p_b=2p_{\bar{b}}$ and $m_\Upsilon=2m_b$. The amplitude of $Z\to \Upsilon(ns)+\gamma$ at the LO is,
\begin{align}
&M_0=\frac{g_W}{2c_W}\frac{e}{3}\frac{1}{(p_b+p_\gamma)^2+m_b^2}\nn\\
&{\rm tr}\left[\Pi\cdot\left(\kappa_Vg_V^b\slashed{\epsilon}(p_Z)-\kappa_Ag_A^b\gamma_5\cdot\slashed{\epsilon}(p_Z)\right)\cdot(m_b-\slashed{p}_b-\slashed{p}_\gamma)\slashed{\epsilon}^*(p_\gamma)\right.\nn\\
&\left.+\Pi\cdot \slashed{\epsilon}^*(p_\gamma)\cdot(m_b+\slashed{p}_b+\slashed{p}_\gamma)\cdot\left(\kappa_Vg_V^b\slashed{\epsilon}(p_Z)-\kappa_Ag_A^b\gamma_5\cdot\slashed{\epsilon}(p_Z)\right)\right].
\end{align}
A simple algebra shows that the partial decay width at the LO is,
\beq
\Gamma_0=\frac{e^2\kappa_A^2(g_A^b)^2}{36\pi m_b}\frac{g_W^2}{c_W^2}\frac{m_Z^4-16m_b^4}{m_Z^5}\Psi_{\Upsilon(ns)}^2(0).
\eeq
It clearly shows that the partial decay width of $Z\to \Upsilon (ns)+\gamma$ will only depend on the  the axial vector component ($\kappa_A$) of the $Zb\bar{b}$ coupling, as expected.
The square of $\Psi_{\Upsilon(ns)}(0)$ in the above equation can be related to the long-distance matrix element introduced in Eq.~\eqref{eq:width} by~\cite{Bodwin:1994jh}
\beq
\Psi_{\Upsilon(ns)}^2(0)=\frac{1}{6N_c}\langle\mathcal{O}^{\Upsilon(ns)}(^3S_1)\rangle.
\eeq
Furthermore, the long-distance matrix element $\langle\mathcal{O}^{\Upsilon(ns)}(^3S_1)\rangle$ can be determined by the partial decay width $\Gamma(\Upsilon(ns)\to e^+e^-)$ in the following numerical analysis; see Sect.~\ref{sec:res}. 
We have checked that our result agrees with that in Ref.~\cite{Huang:2014cxa}. Next, we shall calculate its NLO QCD corrections.

\subsection{The NLO QCD correction}
\begin{figure}
\centering
\includegraphics[scale=0.4]{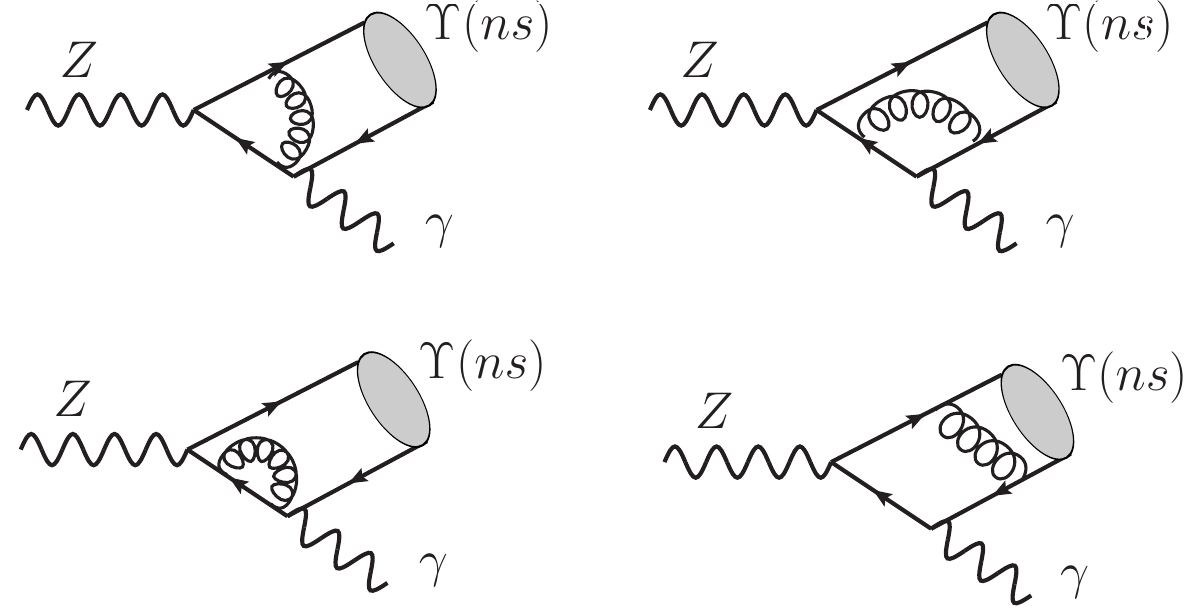}
\caption{Representative one-loop QCD corrections to the first diagram of Fig.~\ref{Fig:FeyLO}.  The QCD correction to the second diagram is similar to this.}
\label{Fig:FeyNLO}
\end{figure}

Some representative one-loop QCD Feynman diagrams for the process $Z\to \Upsilon (ns)+\gamma$ are shown in Fig.~\ref{Fig:FeyNLO}.  The self-energy and triangle diagrams contain the ultraviolet (UV) divergences, while the box diagrams contain the infrared (IR) and Coulomb divergences. To regularize the UV and IR divergences, we adopt the dimensional regularization scheme with $d=4-2\epsilon$. The Coulomb singularity can be isolated by the small relative velocity $v$ between $b$ and $\bar{b}$ quarks. The momenta $p_b$ and $p_{\bar{b}}$ satisfy the relations $\overrightarrow{p}_b+\overrightarrow{p}_{\bar{b}}=\overrightarrow{0}$ and $|\overrightarrow{p}_b-\overrightarrow{p}_{\bar{b}}|=m_bv$~\cite{Kramer:1995nb}. 
The partial decay width  of $Z\to \Upsilon (ns)+\gamma$ at the $\mathcal{O}(\alpha_s)$ can be written as,
\begin{align}
\Gamma_{\rm NLO}&=\Gamma_0\left(1+\frac{\alpha_s}{\pi}C_F\frac{\pi^2}{v}+\frac{\alpha_s}{\pi}F+\mathcal{O}(\alpha_s^2)\right)\nn\\
&\simeq\Gamma_0\left(1+\frac{\alpha_s}{\pi}C_F\frac{\pi^2}{v}\right)\left(1+\frac{\alpha_s}{\pi}F\right),
\end{align}
where the factor $F$, to be explicitly provided in Eq.~\eqref{eq:tot}, is the finite part of NLO QCD correction. The Coulomb singularity arises from the box diagrams, which has been factored out in the above equation and will be absorbed into the long-distance matrix element $\langle\mathcal{O}^{\Upsilon(ns)}(^3S_1)\rangle$ after performing the needed matching procedure in the NRQCD framework. The detailed matching procedure can be found in Ref.~\cite{Bodwin:1994jh} with the replacement of $v\to 2v$ due to the different convention used for its definition.

We should note that the $\gamma_5$ is not well defined in $d$-dimension.  We adopt the Larin scheme~\cite{Larin:1993tq} in this work and express the axial current as 
\beq
\gamma_\mu\gamma_5=Z_5i\frac{1}{6}\epsilon_{\mu\rho\sigma\tau}\gamma^\rho\gamma^\sigma\gamma^\tau.
\label{eq:ga5}
\eeq
The parameter $Z_5$ is the finite renormalization constant, which is introduced to restore the axial current ward identity~\cite{Larin:1993tq}.  At the one-loop level, 
\beq
Z_5=1-\frac{\alpha_s}{\pi}C_F \, ,
\eeq
which yields an additional contribution to the partial decay width as
\beq
\Gamma_{\rm V5}=-2\frac{\alpha_s}{\pi}C_F\Gamma_0.
\eeq

To remove the UV divergences, we choose the on-mass-shell (OS) renormalization scheme in this work. The renormalization constants for the quark field and its mass are, respectively,
\begin{align}
\delta Z_2^{\rm OS}&=-C_F\frac{\alpha_s}{4\pi}\left[\frac{1}{\epsilon_{\rm UV}}+\frac{2}{\epsilon_{\rm IR}}-3\gamma_E+3\ln\frac{4\pi\mu^2}{m_b^2}+4\right],\nn\\
\delta Z_m^{\rm OS}&=-3C_F\frac{\alpha_s}{4\pi}\left[\frac{1}{\epsilon_{\rm UV}}-\gamma_E+\ln\frac{4\pi\mu^2}{m_b^2}+\frac{4}{3}\right],
\end{align}
where $\gamma_E$ is the Euler constant, $1/\epsilon_{\rm UV/IR}$ denote the UV/IR poles, and $\mu$ is the renormalization scale. 
After the renormalization procedure,  all the divergences are cancelled. In the limit of $m_b\to 0$, the contributions to the partial decay width originated from the triangle ($\Gamma_{\rm tri}$), self-energy ($\Gamma_{\rm self}$), counter term ($\Gamma_{\rm CT}$) and box diagrams ($\Gamma_{\rm box}$) are, respectively,  
\begin{align}
\Gamma_{\rm tri}&=\frac{\alpha_s}{\pi}C_F\left[\ln\frac{\mu^2}{m_b^2}+(2\ln2-1)\ln\frac{m_b^2}{m_Z^2}\right.\nn\\
&\left.-\frac{\pi^2}{3}-1+\ln^22+2\ln2\right]\Gamma_0,\\
\Gamma_{\rm self}&=-\frac{\alpha_s}{\pi}C_F\left[\frac{1}{2}\ln\frac{\mu^2}{m_b^2}+\frac{1}{2}\ln\frac{2m_b^2}{m_Z^2}-1\right]\Gamma_0,\\
\Gamma_{\rm CT}&=\frac{\alpha_s}{\pi}C_F\left[-\frac{3}{2}\ln\frac{\mu^2}{m_b^2}+\frac{5}{2}\right]\Gamma_0,\\
\Gamma_{\rm box}&=\frac{\alpha_s}{\pi}C_F\left[\ln\frac{\mu^2}{m_b^2}-\ln2\ln\frac{m_b^2}{m_Z^2}\right.\nn\\
&\left.+\frac{1}{6}\left(\pi^2-30-3\ln^22-12\ln2\right)\right]\Gamma_0.
\end{align}
After matching the calculation in NRQCD with that in full QCD, calculated in the on-shell renormalization scheme, the partial decay width of $Z\to \Upsilon  (ns)+\gamma$, at the NLO in QCD interaction, is found to be 
\begin{align}
\Gamma_{\rm NLO}&=\Gamma_0+\Gamma_{\rm tri}+\Gamma_{\rm self}+\Gamma_{\rm CT}+\Gamma_{\rm box}+\Gamma_{\rm V5}\nn\\
&\simeq \Gamma_0+\frac{\alpha_s(\mu)}{2\pi}C_F\left[(3-2\ln2)\ln\frac{m_Z^2}{m_b^2}\right.\nn\\
&\left.+\ln^22-\ln2-9-\frac{\pi^2}{3}\right]\Gamma_0 \, ,
\label{eq:tot}
\end{align}
where we have explicitly written out the 	
renormalization scale $\mu$ dependence.
The above equation, in the limit of $m_b\to 0$, agrees well with that predicted by applying the LCDA method; see Eq.~(27) of Ref.~\cite{Huang:2014cxa}.
We have also numerically checked that the partial decay width $\Gamma_{\rm NLO}$ in the limit of $m_b\to 0$ approximates well the result with the full $m_b$ corrections included. In the following numerical analysis, we will include the full $m_b$ dependence.

\subsection{Numerical results}
\label{sec:res}
In this work, we adopt the $G_\mu$ scheme~\cite{Dittmaier:2001ay} to fix the electroweak parameters and the SM input values are chosen as follows~\cite{ParticleDataGroup:2020ssz},
\begin{align}
&m_W=80.385~{\rm GeV}, & m_Z&=91.1876~{\rm GeV}, \nn\\
&m_b=4.75~{\rm GeV}, & \Gamma_Z&=2.4952~{\rm GeV}, \nn\\
&G_\mu =1.1663785\times 10^{-5} ~{\rm GeV}^{-2}.
\end{align}
The weak mixing angle is fixed by the ratio $c_W=m_W/m_Z$ and the electromagnetic coupling $\alpha_{\rm EM}=\sqrt{2}G_\mu m_W^2s_W^2/\pi$.

The long-distance matrix element $\langle\mathcal{O}^{\Upsilon(ns)}(^3S_1)\rangle$ can be determined by the partial decay width $\Gamma(\Upsilon(ns)\to e^+e^-)$, which reads as ~\cite{Pineda:1996uk,Pineda:2001et,Beneke:1997jm,Beneke:2014qea,Shen:2015cta}
\beq
\Gamma[\Upsilon(ns)\to e^+e^-]=\frac{2\pi\alpha_{\rm EM}^2}{81m_b^2}\left(1-\frac{16\alpha_s}{3\pi}\right)\langle\mathcal{O}^{\Upsilon(ns)}(^3S_1)\rangle.
\eeq
Based on the experimental measurements of decay branching ratios and total decay widths of $\Upsilon(ns)$~\cite{ParticleDataGroup:2020ssz}, we evaluate  $\langle\mathcal{O}^{\Upsilon(ns)}(^3S_1)\rangle$ at the LO and NLO with strong coupling $\alpha_s=\alpha_s(m_b)$, whose result is shown in Table~\ref{table:PME}.

\begin{table}
\centering
\caption{ Values of the  long-distance matrix elements at the LO and NLO, respectively, for $\Upsilon(ns)$ (in units of ${\rm GeV}^3$), derived from the measurement of partial decay width} $\Gamma[\Upsilon(ns)\to e^+e^-]$.
\begin{tabular}{|c|c|c|c|}
\hline
$\langle\mathcal{O}^{\Upsilon(ns)}(^3S_1)\rangle$&$\Upsilon(1s)$&$\Upsilon(2s)$&$\Upsilon(3s)$\\
\hline
LO& $6.4\pm 0.3$&$3.0\pm 0.4$&$2.2\pm 0.3$\\
\hline
NLO &$10.1\pm 0.5$&$4.8\pm 0.6$&$3.5\pm 0.4$\\
\hline
\end{tabular}
\label{table:PME}
\end{table}

\begin{figure}
\centering
\includegraphics[scale=0.3]{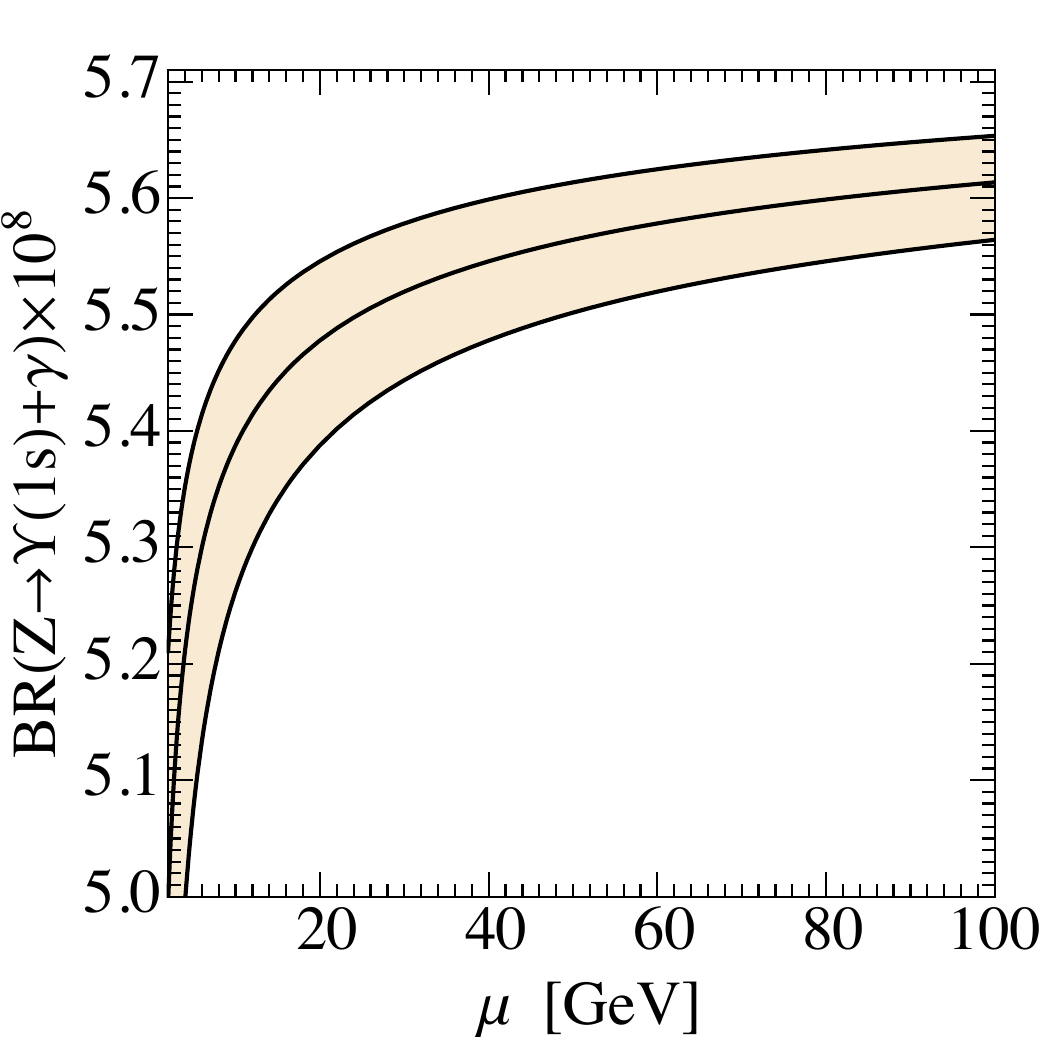}
\caption{The branching fraction of $Z\to \Upsilon(1s)+\gamma$, as a function of the renormalization scale $\mu$, at the NLO with $\kappa_A=1$. The scale uncertainty  is estimated by varying a factor of $1/2$ or 2.}
\label{Fig:BR}
\end{figure}

Figure~\ref{Fig:BR} shows the predicted branching ratio (BR) of $Z\to \Upsilon(1s)+\gamma$ as a function of the renormalization scale $\mu$ at the NLO, with $\kappa_A=1$. Its uncertainty is estimated by varying the scale $\mu$ by a factor of two. The BRs of $\Upsilon(2s)$ and $\Upsilon(3s)$ can be obtained from that of $\Upsilon(1s)$ by simply rescaling their corresponding long-distance matrix elements, cf. Table~\ref{table:PME}. 
In Table~\ref{table:BR}, we compare the BRs of $Z\to \Upsilon(ns)+\gamma$, predicted at the LO and NLO, with $\mu=m_Z$.  
The errors from the scale variation and the long-distance matrix element have been added in quadrature, dominantly determined by the 
latter. Two comments are worth noting.  
Firstly, the $k$ factor from the NLO QCD correction is about 1.5, dominantly arising from the different values of  $\langle\mathcal{O}^{\Upsilon(ns)}(^3S_1)\rangle$ extracted at the LO and the NLO, while the perturbative correction from the one-loop Feynman diagrams is very small.
Secondly, the errors of the long-distance matrix elements (Table~\ref{table:PME}), and consequently, the branching ratios (Table~\ref{table:BR}) presented in this work are larger than those published in Ref.~\cite{Bodwin:2017pzj}. This is because the new experimental data reported in Ref.~\cite{ParticleDataGroup:2020ssz}, which is used in this work, have larger uncertainties than those given in  Refs.~\cite{Chung:2010vz,Bodwin:2017pzj}.

\begin{table}
\centering
\caption{The branching fractions of $Z\to \Upsilon(ns)+\gamma$ at the LO and NLO, respectively, in unites of $10^{-8}$, with the renormalization scale $\mu=m_Z$.} 
\begin{tabular}{|c|c|c|c|}
\hline
${\rm BR}(Z\to \Upsilon(ns)+\gamma)$&$\Upsilon(1s)$&$\Upsilon(2s)$&$\Upsilon(3s)$\\
\hline
LO& $3.83\pm 0.20$&$1.82\pm 0.21$&$1.32\pm 0.17$\\
\hline
NLO &$5.61\pm 0.29$&$2.66\pm 0.31$&$1.93\pm 0.25$\\
\hline
\end{tabular}
\label{table:BR}
\end{table}

\section{The $Zb\bar{b}$ anomalous couplings}

\begin{table}
	\centering
	\caption{Event numbers of the (observed) background and the expected signal, with an assumed branch ratio ${\rm BR}(Z\to \Upsilon(ns)+\gamma)=10^{-6}$, reported in the ATLAS analysis of the $\Upsilon(ns)\to \mu^+\mu^-$ decay channel, at the 13 TeV LHC with an integrated luminosity of $36.1~{\rm fb}^{-1}$~\cite{ATLAS:2018xfc}.} 
	\begin{tabular}{|c|c|c|c|}
		\hline
		Event number&$\Upsilon(1s)$&$\Upsilon(2s)$&$\Upsilon(3s)$\\
		\hline
		Background& 115&106&112\\
		\hline
		Signal &7.8&5.9&7.1\\
		\hline
	\end{tabular}
	\label{table:event}
\end{table}

Owing to the large $Z$ boson production rate at the LHC, the rare decays of $Z\to \Upsilon(ns)+\gamma$ (with $n=1,2,3$) are hopeful to be confirmed at future colliders, and can be used to constrain the $Zb\bar{b}$ anomalous coupling.
Using the inclusive $Z$-boson data sample collected at the 13 TeV LHC, 
with an integrated luminosity of $36.1~{\rm fb}^{-1}$, 
the ATLAS collaboration has obtained 
95\% confidence-level (C.L.) upper limits on the branching fractions of the $Z$ boson decays to $\Upsilon(ns)+\gamma$ of $(2.8,1.7,4.8)\times 10^{-6}$, respectively, via the di-muon decay mode  $\Upsilon(ns)\to \mu^+\mu^-$, assuming Standard Model production~\cite{ATLAS:2018xfc}.
This conclusion is based on the number of observed events and  expected $Z$ boson signals listed in 
Table~\ref{table:event}, where 
the invariant mass of $\mu^+ \mu^- \gamma$ is required to be between 81 GeV and 101 GeV, and ${\rm BR}(Z\to \Upsilon(ns)+\gamma)$ is taken to be $10^{-6}$. 
At the HL-LHC,  a much larger integrated luminosity will be collected, so that one or two orders of magnitude improvement on the upper limits of the BRs, as compared to Ref.~\cite{ATLAS:2018xfc},  could become possible after combining the analyses of the charged lepton decay channels $\Upsilon(ns)\to \ell^+\ell^-$, with $\ell=e,\mu,\tau$, from both the ATLAS and CMS collaborations. 
Working in this scenario, we shall explore the potential of the HL-LHC and future colliders to measuring the $Zb\bar{b}$ anomalous coupling. 
Since the decay branching fractions of $\Upsilon(ns)$ to $e^-e^+$, $\mu^-\mu^+$ and $\tau^-\tau ^+$ channels are about the same~\cite{ParticleDataGroup:2020ssz} and $\tau$-tagging efficiency could reach about $0.6\sim 0.7$~\cite{ATLAS:2015xbi}, we shall  assume in this work the same detection efficiency for all three decay channels, and rescale the event numbers from the 13 TeV ATLAS analysis~\cite{ATLAS:2018xfc} to the 14 TeV HL-LHC  and 100 TeV pp collider.

\begin{figure}
\centering
\includegraphics[scale=0.23]{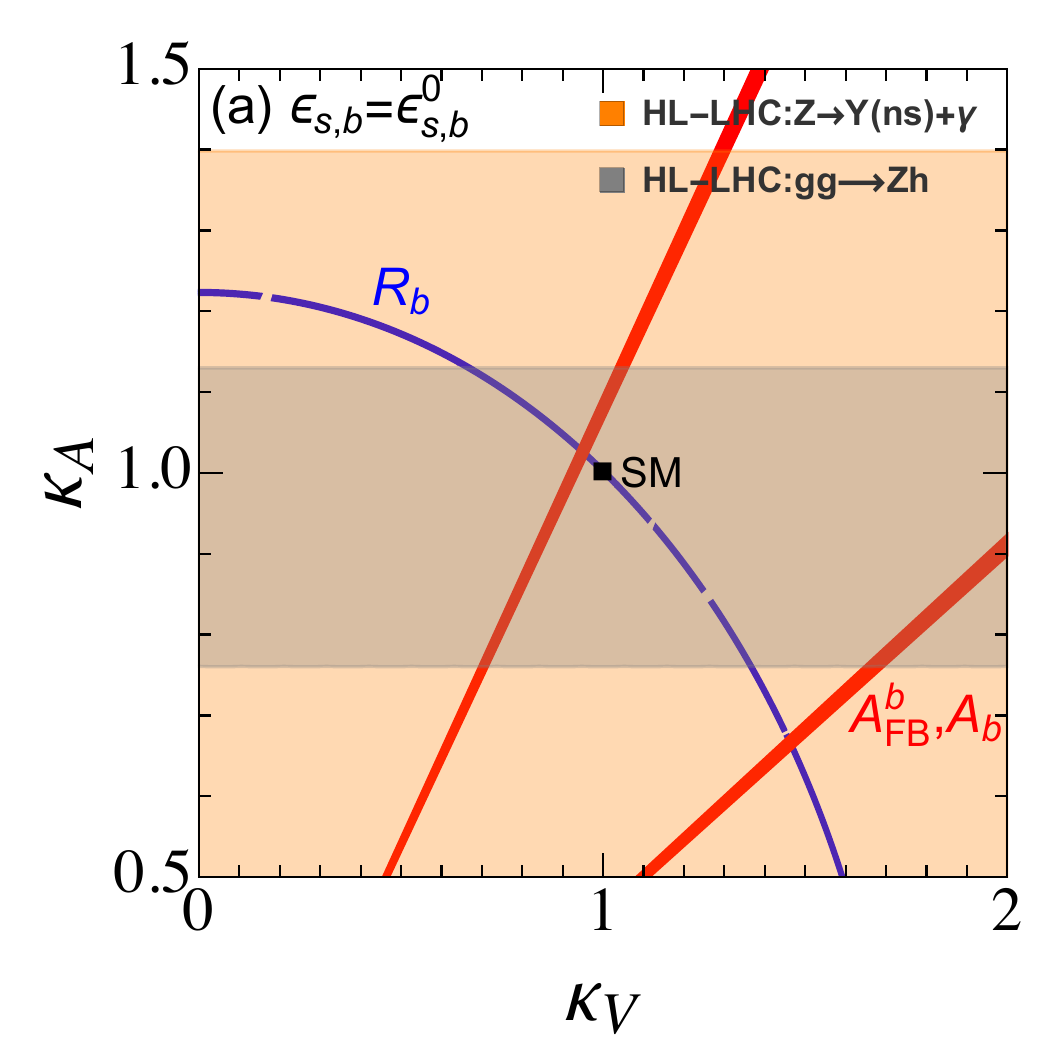}
\includegraphics[scale=0.23]{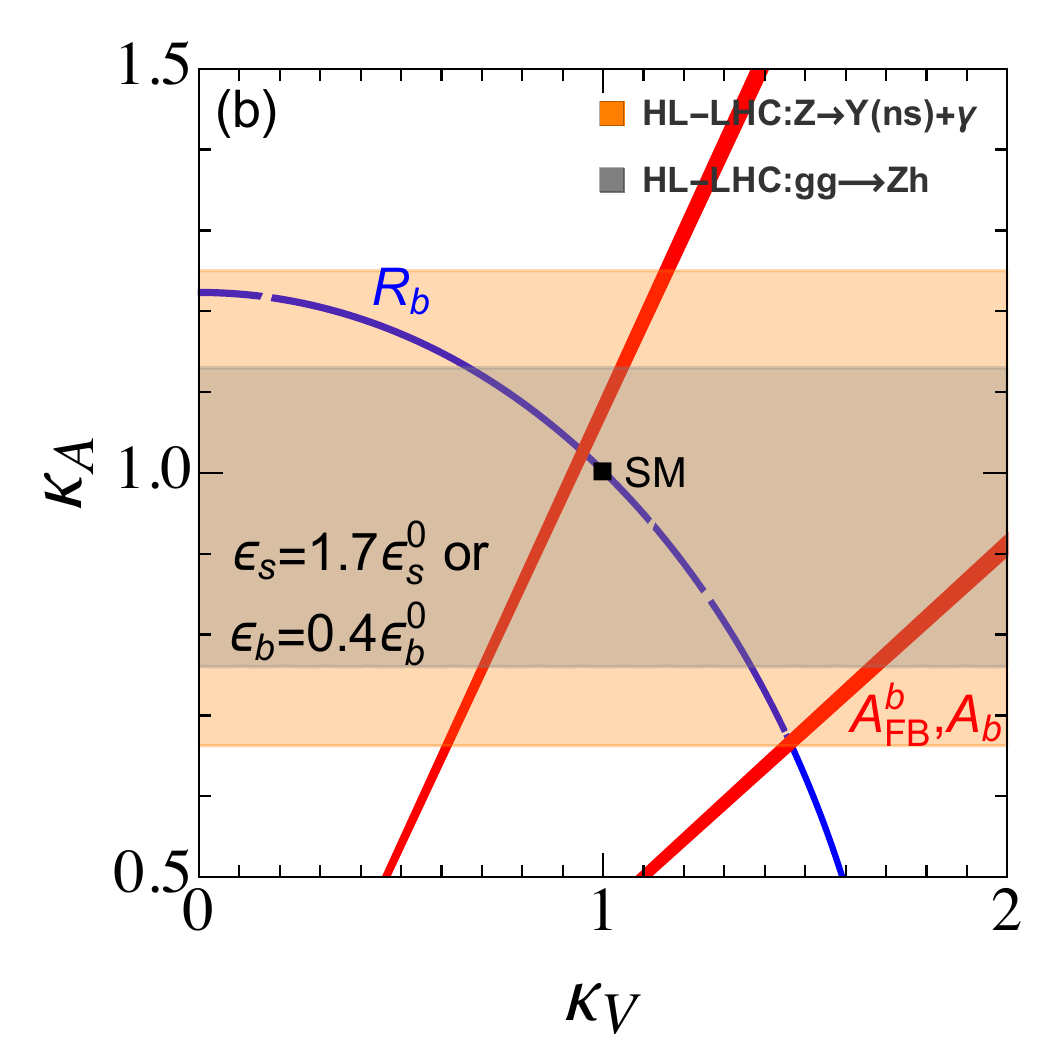}
\includegraphics[scale=0.23]{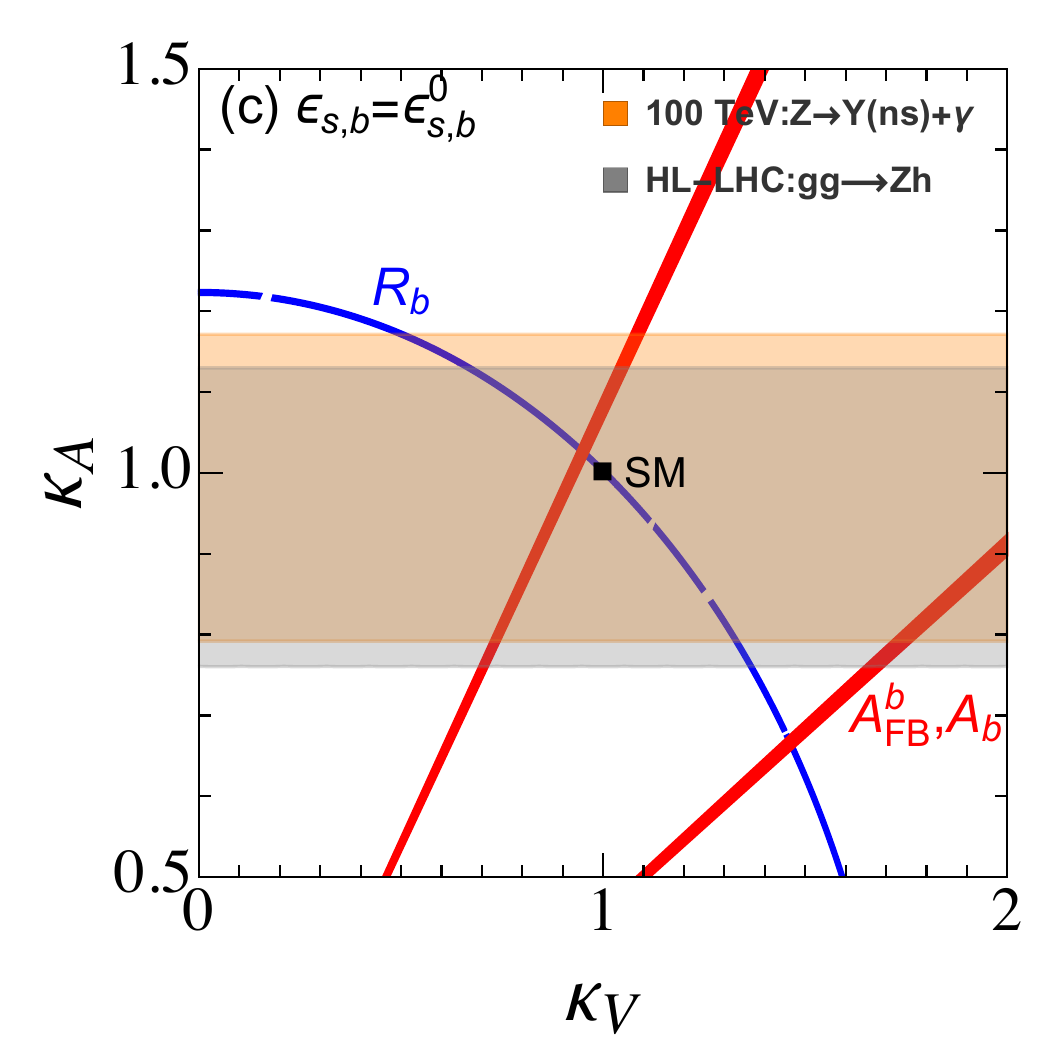}
\includegraphics[scale=0.23]{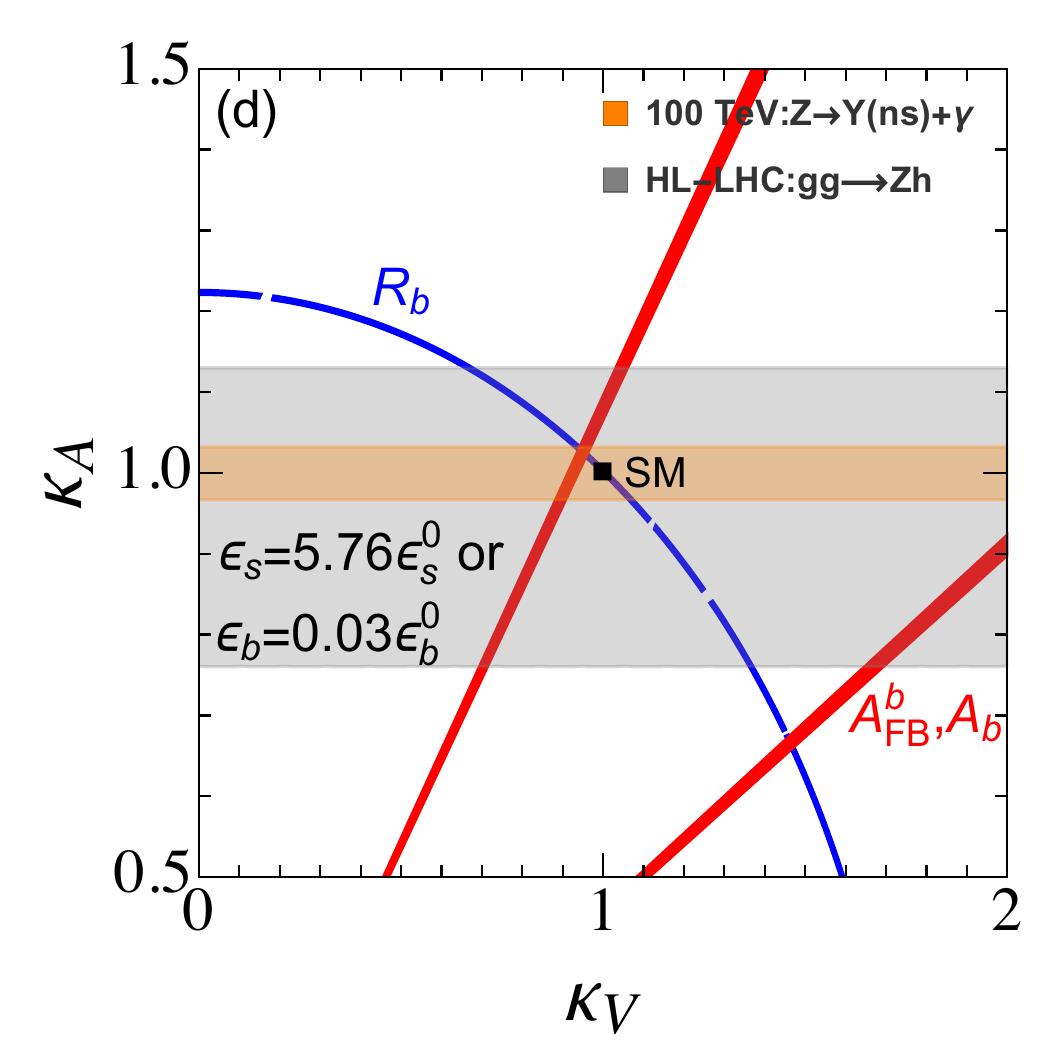}
\caption{The expected 68\% C.L. limits on the $Zb\bar{b}$ anomalous couplings $\kappa_V$ and $\kappa_A$ from the exclusive $Z$ boson decay $Z\to\Upsilon(ns)+\gamma \to\ell^+\ell^-+\gamma$ (orange band) and $gg\to Zh$ scattering (gray band)~\cite{Yan:2021veo}. The blue and red regions come from the $R_b$ and $(A_{\rm FB}^b,A_b)$ measurements at the LEP and SLC, respectively. The parameter $\epsilon_{s,b}^0$ denote the detection efficiencies of the signal and background events, respectively, reported in the ATLAS analysis of $Z\to\Upsilon(ns)+\gamma \to\mu^+\mu^-+\gamma$ at the 13 TeV LHC with an integrated luminosity of $36.1~{\rm fb}^{-1}$~\cite{ATLAS:2018xfc}. }
\label{Fig:zbb}
\end{figure}

To estimate the sensitivity for testing the hypothesis with $\kappa_A\neq 1$ against the hypothesis with $\kappa_A=1$, we define the likelihood function as~\cite{Cowan:2010js},
\beq
L(\kappa_A)=\prod_i\frac{(s_i(\kappa_A)+b_i)^{n_i}}{n_i!}e^{-s_i(\kappa_A)-b_i},
\eeq
where $b_i$ and $n_i$ are the event numbers for the background and observed events in the $i$-th process ($Z\to \Upsilon(ns)+\gamma$ at the ATLAS and CMS collaborations, with $n=1,2,3$ and $\Upsilon(ns)\to \ell^+\ell^-$), respectively. $s_i(\kappa_A)$ is the signal event number with a given value of  $\kappa_A$ for the $i$-th data sample, with $i$ running from 1 to 18 ($=3\times 3 \times 2$) to label data sample for $n=1,2,3$, $\ell=e,\mu,\tau$, and ATLAS or CMS experiments. Here, we assume that the observed event number agrees with the combination of the background and signal events predicted in the SM, {\it i.e.,} $n_i=s_i(\kappa_A=1)+b_i$. With the definition of the test statistic $q$, based on the profile likelihood ratio, as 
\beq
q^2=-2\ln\frac{L(\kappa_A\neq 1)}{L(\kappa_A=1)} \, ,
\eeq
we obtain 
\beq
q^2=2\left[\sum_in_i\ln\frac{n_i}{n_i^\prime}+n_i^\prime-n_i\right] \, ,
\eeq
where $n_i^\prime=s_i(\kappa_A)+b_i$.
An upper limit on the signal strength, as a function of $\kappa_A$, at the 1-$\sigma$ level ({\it i.e.,} 68\% C.L.) corresponds to setting $q=1$.
To apply the above equations, the total event number of the signal ($s_i(\kappa_A)$) and background $(b_i)$ events can be obtained by  properly rescaling the event numbers reported by ATLAS in Table~\ref{table:event}, e.g.
\beq
s_i(\kappa_A)=\kappa^i_{\rm BR}\kappa_{\mathcal{L}}\kappa_{\sigma}s_i^0, \qquad
b_i=\kappa_{\mathcal{L}}\kappa_{\sigma}b_i^0,
\eeq
where $\kappa^i_{\rm BR}=\kappa_A^2 \times {\rm BR_{\rm SM}}^i \times 10^{6}$, and the SM branch ratio ${\rm BR_{\rm SM}^i}$ of $Z\to\Upsilon(ns)+\gamma$ can be found in Table~\ref{table:BR}.
The enhancement factor ($\kappa_{\mathcal{L}}$) arising from the amount of total integrated luminosity is 83. 
The enhancement factor ($\kappa_{\sigma}$)  in the $Z$ boson production cross section at the 14 TeV LHC and 100 TeV pp collider is 1.07 and 7, respectively. 
Though the non-SM $Zb\bar{b}$ couplings could modify the inclusive $Z$ boson cross section predicted by the SM, its contribution is very small (less than a few percent for $\kappa_A \sim 1$) due to the small $b {\bar b}$ parton luminosity and its impact could be ignored in this work.
The result of our analysis is displayed in  Fig.~\ref{Fig:zbb}, where we show the expected 
68\% C.L. constraints on the $Zb\bar{b}$ anomalous coupling obtained from measuring the decay process $Z\to \Upsilon(ns)+\gamma$, produced at the HL-LHC and 100 TeV pp collider.
Fig.~\ref{Fig:zbb}(a) shows that the measurement of the proposed $Z$ boson rare decay production $Z\to \Upsilon(ns)+\gamma$  
at the HL-LHC cannot break the apparent degeneracy of the $Zb\bar{b}$ couplings, found in the electroweak precision measurements, due to the small signal rates. Nevertheless, the total signal efficiency, including the kinematic acceptance, trigger, reconstruction, identification, and isolation efficiencies for the observation of   $\Upsilon(ns)\gamma\to\mu^+\mu^-\gamma$ events produced at the 13 TeV LHC is very low, {\t i.e.,} $\epsilon_s^0=15\%-16\%$~\cite{ATLAS:2018xfc}.
With expected advances in the  experimental measurement and analysis, it is quite possible that  both the signal ($\epsilon_s$) and background ($\epsilon_b$) efficiencies could be improved at the time of HL-LHC runs.
 $\epsilon_{b,s} \equiv \kappa_{b,s}^\epsilon\epsilon_{b,s}^0$, we show the required improvement in the detection efficiencies, with $\epsilon_{s} = 1.7 \epsilon_{s}^0$ {\it or} 
 $\epsilon_{b} = 0.4 \epsilon_{b}^0$,
in order to break the residual degeneracy in the region of $\kappa_{V,A}>0$, {\it i.e.,} to distinguish $(\kappa_V,\kappa_A)=(1.46,0.67)$ from $(0.95,1.03)$. 
Here, we focus on the parameter space with $\kappa_{V,A}>0$, since the off $Z$-pole $A_{FB}^b$ measurements have excluded the region with $\kappa_{V,A}<0$~\cite{Choudhury:2001hs}.
In the same figure, we also show the constraints from the $R_b$ (blue region) and $(A_{\rm FB}^b, A_b)$ (red region) measurements at the $Z$-pole, 
as well as the expected constraint from the measurement of the $gg\to Zh$ production at the HL-LHC (gray region)~\cite{Yan:2021veo}.  To derive  the expected constraint from the $gg\to Zh$ measurement, we have taken into account a factor of 2, as compared to that presented in  Ref.~\cite{Yan:2021veo}, to include both the ATLAS and CMS contributions, which results in a factor of $\sqrt{2}$ reduction in its error band size.
The similar results, but for the 100 TeV pp collider, are shown in Fig.~\ref{Fig:zbb}(c) and (d). 
Even assuming no improvement in the detection efficiencies, with the same $\epsilon_{s}^0$ and $\epsilon_{b}^0$ values found in Ref.~\cite{ATLAS:2018xfc}, the proposed measurements at the 100 TeV pp collider can already break the above-mentioned degeneracy, cf. Fig.~\ref{Fig:zbb}(c), because the inclusive $Z$ boson production cross sections increases by about a factor of 7 as compared to the 14 TeV HL-LHC. However, to exclude the interpretation of the $A_{\rm FB}^b$ data at
LEP by introducing merely the anomalous $Zb\bar{b}$ couplings would require 
$\epsilon_{s} = 5.76 \epsilon_{s}^0$ {\it or} 
$\epsilon_{b} = 0.03 \epsilon_{b}^0$, or some combinations of those two separate improvements, cf. Fig.~\ref{Fig:zbb}(d).

\section{Conclusion}
In this work, we propose to directly measure the axial-vector component of the $Zb\bar{b}$ anomalous coupling by  utilizing the exclusive $Z$ boson rare decay $Z\to \Upsilon(ns)+\gamma$ at the 14 TeV HL-LHC and 100 TeV pp collider. Owing to the $J^{PC}$ quantum number of $ \Upsilon(ns)$ and $\gamma$, we demonstrate that the vector component of $Zb\bar{b}$ coupling can not contribute to the  process $Z\to \Upsilon(ns)+\gamma$. By applying the NRQCD factorization formalism, we calculate the partial decay width  $\Gamma[Z\to \Upsilon(ns)+\gamma]$ to the NLO accuracy in $\alpha_s$. The NLO QCD correction increases the LO decay width by about 50\%, dominantly arising from the difference in the values of the long-distance matrix element $\langle\mathcal{O}^{\Upsilon(ns)}(^3S_1)\rangle$ evaluated at the LO and NLO. The correction from the one-loop Feynman diagrams is quite small. Furthermore, we find a good agreement between the result of  this NLO NRQCD calculation and that of the NLO LCDA calculation~\cite{Huang:2014cxa,Bodwin:2017pzj}, after ignoring the small corrections in powers of  $(m_b/m_Z)^2$ and $v^2$, etc.
To explore the potential of the HL-LHC and the 100 TeV pp collider for constraining the anomalous $Zb\bar{b}$ coupling, we rescale the background and signal event numbers reported in the ATLAS analysis (at the 13 TeV LHC with with an integrated luminosity of $36.1~{\rm fb}^{-1}$)~\cite{ATLAS:2018xfc}. 
Fig.~\ref{Fig:zbb} summaries our findings. It shows that  the HL-LHC can break the degeneracy of the $Zb\bar{b}$ couplings, as implied by the precision electroweak data at LEP and SLC, if the signal  efficiency can be improved by a factor of 1.7 (or the efficiency to suppress the background by a factor of 1/0.4), as compared to the values found in the ATLAS analysis.
At the 100 TeV pp collider, the cross section of inclusive $Z$ boson production could be enhanced by about a factor of 7, so that a better constraint on the $Zb\bar{b}$ coupling is possible. Finally, we note that the observation of the rare decays $Z\to \Upsilon(ns)+\gamma$ at hadron colliders provides complementary information to the jet-charge weighted single-spin asymmetry measurement at the EIC and the $gg\to Zh$ production rate measurement at the LHC for determining the $Zb\bar{b}$ couplings.

Note Added:  After the completion of this work, we were asked to comment on the production rate of $Z\to \chi_b+\gamma$. We found that its branching ratio is around $\mathcal{O}(10^{-10}-10^{-9})$, which is more than one order of magnitude smaller than the BR of $Z\to \Upsilon(ns)+\gamma$.

\vspace{3mm}
\medskip
\noindent{\bf Acknowledgments.}
The work of P. Sun is supported by Natural Science Foundation of China under grant No. 11975127 and No. 12061131006 as well as Jiangsu Specially Appointed Professor Program.
BY is supported by the U.S. Department of Energy, Office of Science,
Office of Nuclear Physics, under Contract DE-AC52-06NA25396 through the LANL/LDRD Program, as well as the TMD topical collaboration for nuclear theory. CPY is supported by the U.S.~National Science Foundation under Grant No.~PHY-2013791. C.-P.~Yuan is also grateful for the support from the Wu-Ki Tung endowed chair in particle physics.

\bibliographystyle{apsrev}
\bibliography{reference}

\end{document}